# Probing confined interfacial excitations in buried layers by Brillouin light scattering

X. Zhang, R. Sooryakumar

Department of Physics The Ohio State University  Columbus, OH 43210.

Brillouin light scattering from silicon oxynitride films grown on GaAs reveals a low frequency elastic wave excitation at frequencies lying below that of the Rayleigh surface wave.  This mode, identified as an excitation localized by the interface, arises from the presence of a soft, thin transition layer between the film and substrate. The results are discussed in the framework of a Green's function formalism that reproduces the experimental features and illustrates the nature of the mode and its difference from Stoneley excitations that are predicted in special cases for an abrupt interface separating two semi-infinite solid media. Observations of this low frequency excitation offer a previously unexplored approach to characterize, non-destructively, the elastic properties of buried interfaces.

Characterization of the interface separating two layers remains a challenging problem in the field of thin film/ multi-layer growth and in applications relying on high quality microstructures. While the interface in such structures is important in determining the quality of the films, the transition from one material layer to the other is, in most cases, not abrupt. The properties of such transition layers are, in turn, important for they often act as a seed for the subsequent growth and therefore play a significant role in defining the properties associated with the ensuing film. For instance, the elastic properties of the intermediary layer are important for they provide the bridge that relates the lattice parameters of the two adjacent layers and thus the strain and structural quality of the film being deposited. The most common technique utilized in probing this buried region is transmission or scanning electron microscopy (SEM) which, despite its high resolution, is burdened by being a destructive method. Amongst non-destructive techniques, inelastic laser light scattering has had very limited success in probing the interfacial region. This method relies on the detection of Stoneley acoustic waves[1] that are guided along the abrupt interface of two semi-infinite solid media in contact along a plane.  The biggest drawback of exploiting Stoneley waves to probe the interfacial properties is that strict restrictive conditions on the acoustic parameters defining the two media immediately bordering the abrupt interface must be met.[2,3]  In particular, the Stoneley wave exists *only* when the shear acoustic velocities of the two adjacent layers are comparable. Thus the technique has been effective only in very few layered material combinations; for example in molybdenum films on fused.[4]

In this paper we report on the experimental observation and provide theoretical insight into the existence of an acoustic excitation localized by the interfacial region in the case where the two layers are separated by a thin transition layer. Brillouin light scattering (BLS) reveals this mode as a well defined low frequency excitation while simulations based on a surface Green's function formalism shows that its displacement amplitude falls off exponentially to zero in both the film and substrate with the excitation being guided by the very thin transition layer. Thus while the mode is indeed localized by the interface, unlike the Stoneley excitation, it is not encumbered by stringent conditions on the shear velocities of the two layers. Therefore observation and investigation of this excitation provides a previously unexplored avenue to non-destructively characterize the properties of buried interfacial layers of hetero-structures.

The silicon oxynitride films used in this study were grown to thickness' $d = 1$, 2 and $3 \mu m$ on (001) GaAs wafers as discussed elsewhere[5]. The BLS measurements were performed in back-scattering at ambient temperature using approximately 70 mW p-polarized 514.5 nm radiation. The instrumental resolution of the measurement is 0.1 GHz. The guided interfacial excitation is observed at low frequencies lying even below the Rayleigh wave and scattering angles $\theta > 40°$, together with suitable slits, were used to reduce broadening of the surface mode peaks.

Fig. 1(a) shows low frequency unpolarized Brillouin spectra ($p \rightarrow p$ and $p \rightarrow s$ scattering) recorded along the [100] direction from the $1 \mu m$ and $2 \mu m$ thick films at a scattering angle of $\theta = 70°$ while those from the $3 \mu m$ film were at angles ranging from $\theta = 40°$ to $\theta = 65°$. The main aspect of the data is contained in the low frequency spectra that appear as a closely spaced triplet-structure whose peak frequencies extend from about 11 GHz to 13 GHz for $\theta = 70°$. The highest frequency mode that appears in cross-polarization ($p \rightarrow s$) has been previously identified as a shear horizontal resonance (SHR).[6] The surface Rayleigh excitation lies just below the SHR and its comparitive weakness is due to the relatively low refractive index ($n = 1.59$) of the oxynitride film and resulting weak reflection from the free surface. The elastic constants ($C^{film}_{11} = 95.9$ GPa, $C^{film}_{44} = 33.6$ GPa) for these films were deduced from the dispersion of the SHR,[6], longitudinal resonance[6] and organ-pipe type standing waves.[7] The lowest frequency excitation of this triplet of modes is the focus of this study.

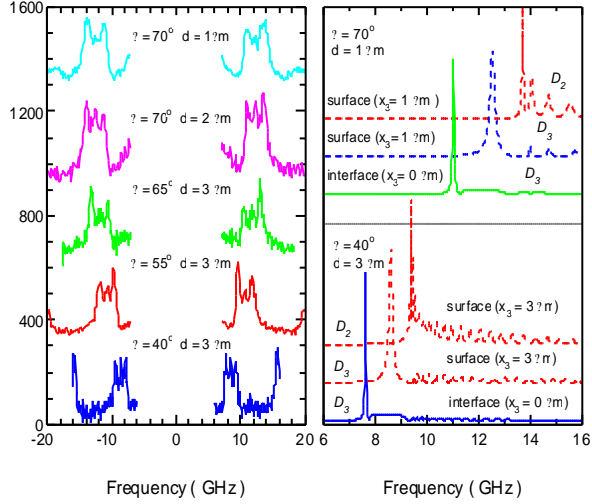

Fig. 1 (a). Brillouin light scattering spectra for silicon oxynitride films on GaAs substrate. The peak with lowest frequency is identified as a buried interfacial mode. The intermediate and high frequency modes are respectively the Rayleigh wave and shear horizontal resonance. (b) Calculated LDOS for the modes at the free surface ($D_2$ and $D_3$, $x_3 = d$) and interface ($D_3$, $x_3 = 0$) for 1 μm thick film at incident angle of $70°$ and for 3 μm sample at incident angle of $40°$ with a transition layer thickness of $l = 40$ nm and $C = 0.37$.

In order to gain insight into the origin and properties of the low lying excitations, the projected local density of states (LDOS) $D_i$ was evaluated within the Green's function formalism[8,9]

$$D_i(\omega^2, K, x_3) = -(\pi)^{-1} \text{Im } G_{ii}(K, x_3, \omega^2).$$

Here $i$ refers to the mode polarization; $i = 1$ for longitudinally polarized excitation, 2 and 3 respectively, for shear horizontal and sagittal polarization normal to the surface. $G_{ii}$ is the $(x_i, x_i)$ component of the Fourier (frequency and wavevector) domain elastodynamic Green's function tensor for depth $x_3$. The method of calculation of $G_{ii}$ is provided in reference 10.

The spatial distribution of the mode displacements is evaluated through the relation[10]

$$U_j(k_{//}, x_3; \omega) = \sum_{n=1}^{m} \frac{i}{\omega}(B^{-1})_j^{(n)} U_j^{(n)} \exp\{-i\, k_3^{(n)} x_3\}$$

where j identifies the displacement components, m the number of eigen modes, $U^{(n)}_j$ the vector components of the eigen mode, B the boundary condition determinant and $k^{(n)}_3$ the wave vector component of the mode along $x_3$, the film normal.

In order to support a mode with a frequency lower than the transverse wave thresholds of the film and substrate, we postulate the existence of a soft transition layer between the film and substrate. The presence of such a thin intervening layer is possible during initial stages of growth of a supported film. In fact, SEM imaging of the interfacial region confirms the existence of such a narrow transition layer. The inset to figure 4 illustrates an example of the SEM image recorded in the proximity of the substrate-film interface and reveals the existence of a 40nm thick transition layer in the 3 μm film. In order to model such a structure, we consider the $l = 40$ nm thick transition layer to have the same mass density as the oxynitride film while its elastic constants are taken as a variable lying below the $C_{ij}$'s of the film. We consider the parameter $C = C_{44}^{soft}/C_{44}^{film}$, the ratio of the shear elastic constant of the soft transition layer to that of the oxynitride film to be a fitting parameter.

Figure 1(b) shows the calculated local density of states for the modes at the film surface ($D_2$ and $D_3$, dashed lines) at $x_3 = d$, and the buried film-substrate interface ($D_3$, solid line) at $x_3 = 0$ for two samples having thickness $d = 3$ μm and 1 μm. The thickness $l$ of the intervening layer in each case is taken to be 40 nm and, as discussed below, the softness parameter $C$ ($= C_{44}^{soft}/C_{44}^{film}$) is equal to 0.37. In this case the calculated LDOS for the 3 μm film ($\theta = 40°$) and the 1 μm film ($\theta = 70°$) can be directly compared to the experimental spectra illustrated in figure 1(a). In each case the density of states have been displaced for clarity. It is evident that the three main peaks in the recorded spectra (Fig. 1a) are well reproduced in figure 1(b). The two higher lying modes within the triplet structure of Fig. 1(a) are modes associated with the large density of states ($D_2$ and $D_3$) in the vicinity of the upper surface, while the lowest frequency mode is characterized by an enhanced mode density ($D_3$) lying close to the interface. This behavior is in agreement with the previous assignment[6] of the two higher modes to the Rayleigh and SHR excitations that are primarily localized to the upper surface. As $\theta$ increases from $40°$ to $65°$ the three peaks in the calculated LDOS from the three films move (not shown in Fig. 1(b)), in close agreement with observation, to higher frequencies.

We now consider, and dismiss, the possibility that the lowest frequency mode (Fig. 1a) is a Stoneley wave localized to an abrupt film-substrate interface. The absence of a true Stoneley wave in the oxynitride-GaAs structure simply follows from the strict, narrow range of material parameters (density $\rho$ and elastic constants) required, but not satisfied in the present case, wherein the shear velocities of the two media must be comparable [$\{C^{film}_{44}/C^{GaAs}_{44}\} \sim \{\rho^{film}/\rho^{GaAs}\}$].[2,3] This is further confirmed by the mode density calculations shown in figure 2 for $C = 1$ that relates to an abrupt interface. In this case with no intermediate transition layer, a broad shoulder in the density of states occurs and spans between the transverse wave threshold (11.3 GHz) in the substrate ($\omega_l^S$) to the corresponding threshold $\omega_l^F$ (13.7 GHz) in the film. The weak maximum evident at 12.5 GHz between $\omega_l^S$ and $\omega_l^F$ corresponds to the frequency of the fast transverse wave in the substrate. In general, waves with a frequency lower than the threshold frequency become evanescent. Thus excitations lying in the frequency range between $\omega_l^S$ and $\omega_l^F$ are evanescent in the film but show oscillatory behavior within the substrate with a reduced amplitude well within the substrate. Such evanescent and oscillatory behavior for C =1 of the mode amplitude $U_3$ in the film and substrate is illustrated in figure 3(a) for a mode frequency of 12.5 GHz, lying between the two thresholds. This type of wave is leaky[11,12] since energy is drawn from the interface to the substrate. Because of their lower LDOS, such waves are generally not observed by light scattering.

Figure 2 shows the influence of varying C, the relative softness of the transition layer, on the mode density at the buried interface. As C decreases from 1 to 0.3 for a constant transition layer thickness of l = 40 nm, the emergence of a relatively broad, asymmetric resonance peak that moves to lower frequency and appears as a shoulder near $\omega_l^S$ is evident. When C = 0.37, the peak has crossed to frequencies below $\omega_l^S$ and is close to 11 GHz. The latter corresponds to the low frequency peak observed from this d = 3 µm film in the BLS experiment.

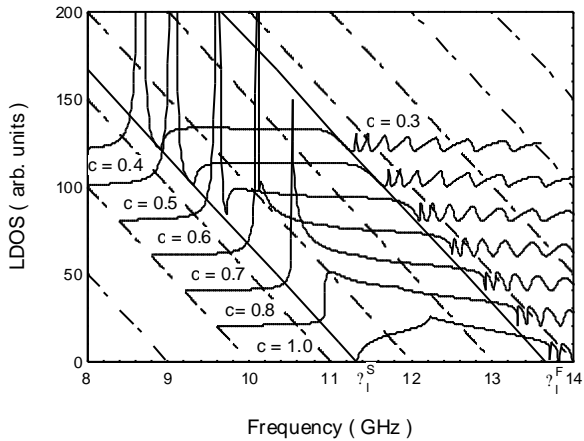

Fig. 2. The influence of varying the parameter C, the relative softness of the transition layer, on the mode density at the buried interface for the 3 µm thick film. The thickness l of the soft layer is 40 nm. $\omega_l^S$ (11.3 GHz) and $\omega_l^F$ (13.7) are transverse wave threshold for the substrate and film respectively. When the ratio C is small enough, a sharp peak emerges and moves to frequencies lower than $\omega_l^S$. C = 1 corresponds to an abrupt film-substrate interface.

In order to analyze the properties of this low frequency excitation, its spatial displacement distribution is illustrated in figure 3(b). The distribution is calculated for a value of C = 0.37, interfacial layer thickness of l = 40 nm and a frequency of 11.0 GHz for the d = 3 µm thick film. The displacement amplitude $U_3$ decreases in both the film and the substrate and eventually vanishes when far removed from the region of the interface. Due to the narrow width (40 nm) of the transition layer, the maximum displacement $U_3^2$ appears in the substrate. For larger values of l, the maximum moves into the buried layer. The spatial distribution of the new mode does not necessarily decay with distance within the transition layer and could oscillate prior to emerging into the substrate or film. This follows from the fact that the transverse wave frequency threshold ( 8.31 GHz) of the soft layer ($C^{soft}_{44} = 0.37\ C^{film}_{44}$), is lower than the mode frequency (11.0 GHz).

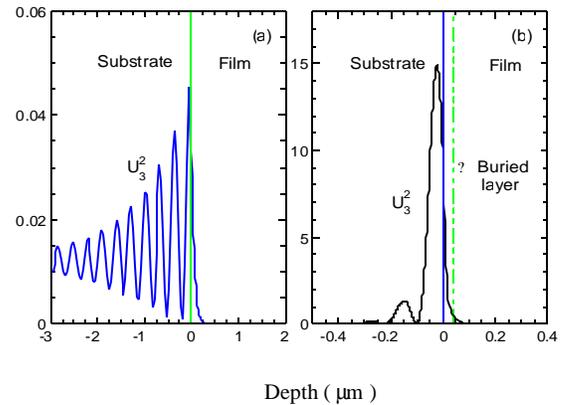

Depth ( µm )

Fig. 3. The spatial displacement distribution a) for the leaky interfacial mode with frequency of 12.5 GHz in the absence of a transition layer. b) for the first order interfacial mode with frequency of 11 GHz when transition layer thickness l = 40 nm, ratio C = 0.37.

As discussed, calculations such as those associated with figure 1(b) account well for the observed lowest frequency peak for all samples and its dependence on scattering angles. This confirms the interfacial aspect of the mode. Therefore it is the presence of a soft transition layer separating the film from the substrate that provides the existence criterion for the interfacial mode. This layer should be elastically soft to reduce the mode frequency to lie below both transverse wave thresholds for the film and substrate. This results in not only the mode being localized at the film-substrate interface but also yields a large LDOS at the interface (Fig. 1(b)) which enables its observation via light scattering.

Figure 4 illustrates the role of the thickness (l ) and elastic properties (C) of the buried transition layer in determining the stability and frequency of the interfacial wave. The solid curve in Fig 4 depicts this mutual relationship between l and C for the interfacial wave (at ω = 11.0 GHz) for the present film-substrate sample and scattering geometry. It is seen that when l decreases, the elastic parameter C must also diminish to stabilize the interfacial mode. For example a thickness of l = 40nm, the transition layer thickness in the present samples, the elastic constant of the layer should be about 37% of film elastic constant (C = 0.37) for the appearance of the mode at 11 GHz. In fact, the onset of the interfacial mode occurs immediately after its frequency falls below the substrate transverse wave thresholds $\omega_1^S$. The dashed curve in figure 4 identifies this onset with the interfacial excitations stabilizing for transition layer material parameters lying below this curve. It is evident that the thicker the transition layer, or the smaller the C parameter, the lower the interfacial mode frequency.

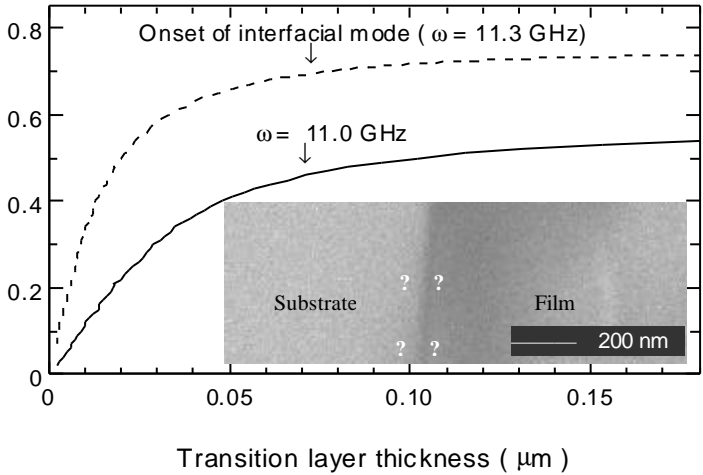

Fig. 4. The relation connecting the transition layer thickness l and elastic constant ratio C for existence of the interfacial mode (solid curve). The dashed curve corresponds to the onset of the interfacial mode at a frequency of 11.3 GHz, which is the transverse wave frequency threshold $\omega_1^S$ of the substrate. The inset to the figure shows the SEM image in the vicinity of the film-substrate interface for the d = 3μm film and reveals the presence of a thin transition layer of thickness l = 40 nm.

The 11 GHz mode observed in our study is in fact the first of a series of higher order interfacial excitations when the thickness of the transition layer is large or when its shear elastic constants are small. In such a case the first order interfacial mode softens, and is accompanied by the occurrence of high order excitations. This behavior is similar to the presence of high order Sezawa[10] or Lamb waves[13,14] in thin supported films. The displacement of the predicted higher order modes when the transition layer become softer and wider may decay or display oscillatory behavior in the substrate or film prior to eventually decaying exponentially at distances well away from the soft intervening layer.

In conclusion, an interface localized mode has been successfully observed by Brillouin light scattering in oxynitride-GaAs structures and its properties simulated by a surface Green's function technique. The appearance of the mode indicates the presence of a soft transition layer between the film and substrate which helps to lower the frequency of the mode and results in a large LDOS for the mode that enables its observation through light scattering. The amplitude of the mode displacements falls off exponentially to zero on the both sides of the interface. The mode is localized close to interface and does not require, as in the case of Stoneley waves, strict conditions of the film and substrate material parameters for its existence. Its relative ease of observation via light scattering experiments provides a *non-destructive* approach to characterize the elastic properties of buried films and transition layers.

We are grateful to T.S. Hickernell for providing the samples. Work at The Ohio State University was supported by the Army Research Office under grant DAAD 19-00-1-0396 and the National Science Foundation under grant 97-01685.